\documentstyle[12pt]{article}
\begin{document}
\begin{flushright} UP--TP 94/1\end{flushright}
\sloppy
\begin{center}
{\bf TIME-LIKE SHOCK HADRONIZATION \\ OF A SUPERCOOLED QUARK-GLUON
PLASMA}
\\[1cm]
{\bf M.I. Gorenstein}, \footnote{Permanent
address: Institute for Theoretical Physics, 252143 Kiev-143, Ukraine}
{\bf H.G. Miller}, {\bf R.M. Quick} and {\bf R.A. Ritchie}
\\[1cm]
Department of Physics, University of Pretoria,  Pretoria 0002,\\
South Africa
\\[3cm]
\end{center}

\begin{abstract}
We study the energy-momentum and baryonic number conservation
laws for quark-gluon plasma discontinuity transitions into hadron
matter states. We find that the time-like shock hadronization
of a supercooled quark-gluon plasma
(when the normal vector to the discontinuity hypersurface is time-
like)
should take place. We consider some properties of this process, which
is different from the standard space-like shock hadronization.

\end{abstract}
\newpage

Relativistic shock phenomena have been widely discussed in recent
years
within their connection to high-energy heavy ion collisions (see,
for example, [1]).
Special attention has been paid to the phase transition between the
hadronic matter (HM) and a quark-gluon plasma (QGP) expected
to occur in such collisions. Different scenarios involving shocks as a
possible mechanism for both the deconfinement transition and for QGP
hadronization have been suggested.
The only physical requirements which regulate the dynamics of these
shock transitions are the energy-momentum and baryonic number
conservation
across the discontinuity front. The entropy growth condition
(thermodynamical stability) and mechanical stability conditions [2]
should be additionally checked to guarantee shock existence.

In the compressional
shock model of heavy ion collisions the discontinuity
in the thermodynamical parameters of
strongly interacting matter appears
due to the initial hydrodynamical (velocity) discontinuity in the
system
of colliding nuclei. The physical picture of (one-dimensional)
compression
shocks in heavy ion collisions is quite transparent (see, for
example, [3]).
It does not mean
of course that the assumptions of this model are definitely realized
in
the real world -- only experimental studies and theoretical analysis
will
clear up this question.

The status of the QGP shock-like hadronization, suggested in Ref.[4],
is much less clear.
Firstly, the origin of a hadronization
discontinuity is quite different. It can result
from  smooth initial conditions in the course of the QGP expansion
if a transition between HM and QGP is a 1-st order phase transition.
In this case one can also expect the appearance of supercooled QGP
states ahead of the discontinuity front and superheated HM states
behind it.
Secondly, the space-time properties of a discontinuity
hypersurface as well as the physical meaning of the collective
velocities have not yet been studied in this case.
A recent consideration of shock-like QGP hadronization can be found
in Refs.[5].

In our paper we analyze the
discontinuity-like hadronization model. By admitting the existence of
supercooled QGP and superheated HM we have the picture of
discontinuity-like transitions which is essentially richer than only
standard compression and rarefaction shocks (see also [6,7]). We find
that
discontinuity hadronization across the hypersurface with a time-like
normal 4-vector (we call them t.l. shocks) should occur
in the QGP dynamical evolution (i.e., in its expansion process).
The mathematical properties of these t.l. shocks
are in some respects rather similar to that of standard
shock waves (we call them space-like shocks, or s.l. shocks),
considered
in Refs.[5],
for which the normal 4-vector
to shock front hypersurface is always space-like. Both kinds of
shocks satisfy the same shock adiabat equation (Taub adiabat).
The physical picture of t.l. schocks and its consequences
for QGP hadronization are, however, quite
different and will be studied in our paper.

The system evolution in relativistic hydrodynamics is governed by
the energy-momentum tensor
$ T^{\mu\nu} = (\epsilon + p)u^{\mu}u^{\nu} - pg^{\mu\nu}$
and conserved charge currents
(in our applications to heavy ion collisions it is the baryonic
current
$ nu^{\mu}$).
They consist of local thermodynamical fluid quantities
(the energy density $\epsilon$,
pressure $p$, baryonic density $n$)
and the collective four-velocity
$u^{\mu} = (1-{\bf v}^{2})^{-1/2}(1,{\bf v})$.
The continuous flows are the solutions of the
hydrodynamical equations
$\; \partial T^{\mu\nu}/ \partial x^{\mu} = 0\;$,
$\; \partial nu^{\mu}/ \partial x^{\mu} = 0\;$
with specified initial and boundary conditions.
These equations are nothing more than the differential form
of the energy-momentum
and baryonic number conservation laws. Along with these continuous
flows,
the conservation laws can also be realized in the form of
discontinuous
hydrodynamical flows which are called shock waves and satisfy the
following equations:
\begin{equation}
 T_o^{\mu\nu}\Lambda_{\nu} =
 T^{\mu\nu}\Lambda_{\nu}  \;\; ,
\end{equation}
\begin{equation}
 n_{o}u_{o}^{\mu}\Lambda_{\mu}=
 n u^{\mu}\Lambda_{\mu} \;\; ,
\end{equation}
where $\Lambda ^{\mu}$ is the unit 4-vector normal to the
discontinuity
hypersurface. In Eqs.(1,2) the
zero index corresponds to the initial state ahead of
the shock front and quantities without an index are the final state
values
behind it. A general derivation of the shock equations (valid for both
space-like and time-like normal vectors $\Lambda ^{\mu}$) was given
by L.Csernai [6].

The important constraint on the transitions (1,2)
(thermodynamical stability condition) is the requirement of non-
decreasing
entropy ($s$ is the entropy density):
\begin{equation}
su^{\mu}\Lambda _{\mu} \ge s_{o}u_{o}^{\mu}\Lambda _{\mu} \; .
\end{equation}

To simplify our consideration and make our arguments more transparent
we consider only one-dimensional hydrodynamical motion in what
follows.
To study the s.l. shock transitions one can always choose
the Lorentz frame
where the shock front is at rest. Then the hypersurface of
shock discontinuity is
$x_{s.l.sh} = const$, and $\Lambda ^{\mu} =(0,1)$. The shock equations
(1,2) in this (standard) case are:
\begin{equation}
T_{o}^{01} = T^{01}, \; \; T_{o}^{11} = T^{11} \;\; ,
\end{equation}
\begin{equation}
n_{o}u_{o}^{1} = nu^{1} \; \; .
\end{equation}
Solving Eq.(4)
one obtains
\begin{equation}
v_{o}^{2} = \frac{(p - p_{o})(\epsilon + p_{o})}
{(\epsilon - \epsilon_{o})(\epsilon_{o} + p)} \; , \; \;
v^{2} = \frac{(p - p_{o})(\epsilon_{o} + p)}
{(\epsilon - \epsilon_{o})(\epsilon + p_{o})} \;\; .
\end{equation}
Substituting (6) into (5) we obtain the well known
Taub adiabat equation (TA) [8]
\begin{equation}
n^{2}X^{2} - n_{o}^{2}X_{o}^{2} - (p - p_{o})
(X + X_{o}) = 0 \;\; ,
\end{equation}
in which $X \equiv (\epsilon + p)/n^{2}$, and it therefore
contains only thermodynamical variables.

For discontinuities on a hypersurface with a time-like normal vector
$\Lambda^{\mu}$ one can
always choose another convenient Lorentz frame (``simultaneous
system'') where
the hypersurface of the discontinuity is
$t_{t.l.sh} = const$ and $\Lambda^{\nu}
= (1,0)$. Equations (1,2) are then
\begin{equation}
T_{o}^{00} = T^{00}, \; \; T_{o}^{10} = T^{10} \;\; ,
\end{equation}
\begin{equation}
n_{o}u_{o}^{0} = nu^{0} \;\; .
\end{equation}
Solving Eq.(8) we find
\begin{equation}
\tilde{v}_{o}^{2} =
\frac{(\epsilon - \epsilon_{o})(\epsilon_{o} + p)}
{(p - p_{o})(\epsilon + p_{o})}, \; \;
\tilde{v}^{2} = \frac
{(\epsilon - \epsilon_{o})(\epsilon + p_{o})}
{(p - p_{o})(\epsilon_{o} + p)} \;\;,
\end{equation}
where we use $``\sim"$ sign to distinguish the t.l. shock case
(10) from the standard s.l. shocks of (6). Substituting (10) into (9)
one
finds the equation for t.l. shocks  which is identical to
the TA  of Eq.(7).
We stress that the intermediate steps (Eqs.(10)
and (6)) are, however, quite different.
Note, that the two solutions, Eqs.(10) and (6), are connected to
each other by simple relations
\begin{equation}
\tilde{v}_{o}^{2} =
\frac{1}{v_{o}^{2}}, \; \;
\tilde{v}^{2} = \frac
{1}{v^{2}} \;\; ,
\end{equation}
between velocities for s.l. shocks and t.l. shocks.
These relations show that only one kind of
transition can be realized for a given initial state and final state.
Physical regions $[0,1)$ for $v_{o}^{2},v^{2}$ (6) and for
$\tilde{v}_{o}^{2},\tilde{v}^{2}$ (10) can be easily found in
$(\epsilon $--$p)$-plane.
For a given initial state $(\epsilon _{o},p_{o})$ they are shown
in Fig.1.
If one takes as initial and final states only states which are
thermodynamically equilibrated
it can be proven that the physical conditions
$0 \leq v_{o}^{2}, \;
v^{2} \leq 1 \;\;$
are satisfied only for s.l. shocks for any EOS
which gives the speed of sound in the medium
smaller than or equal to 1.
The TA passes then through the point $(\epsilon _{o},p_{o})$ (which
is called
the centre of TA) and lies as a whole in the regions I and IV in
Fig.1. The
compressional s.l. shock transitions into  region IV
and rarefaction s.l. shock transitions into
region I are called detonation and deflagration respectively.
For supercooled initial QGP states the TA no longer passes through
the point
$(\epsilon _{o},p_{o})$ and
new possibilities of t.l. shock hadronization ((8,9) shock transitions
to regions III and VI in Fig. 1) appear.

To study any hydrodynamical problem quantitatively we need one more
equation -- an
equation of state (EOS) which expresses local thermodynamical
parameters in terms of two independent variables. The most convenient
way is to present it in the form of $p=p(T,\mu)$ with temperature $T$
and baryonic chemical potential $\mu$ as independent variables.
All other thermodynamical functions can be found then from
the following thermodynamical identities:
\begin{equation}
s = \left (\frac{\partial p}{\partial T}\right )_{\mu}\;,\;\;
n = \left (\frac {\partial p}{\partial\mu}\right )_{T}\;,
\;\; \epsilon = Ts + \mu n - p\;.
\end{equation}
For HM we use a thermodynamically consistent
``excluded volume" model [9] where the hadronic pressure can
be expressed in terms of ideal gas pressures as ($V^{*}_{i}$ is the
proper volume for a hadron of type $``i"$):
\begin{equation}
p_{h}= \sum_{i} p_{i}^{id}(T,\tilde{\mu _{i}})\;,\;\;\;
\tilde{\mu _{i}}=\mu _{i}-V^{*}_{i}p_{h}\;.
\end{equation}
The hadron pressure is
suppressed compared to the ideal gas pressure because
of the shift, $-V^*_i p_{h}$, in the particle chemical potentials.
In  the entropy, baryonic
 and energy density, as follows from Eqs.(12), the additional
suppression
factor $[1+\sum_{i} V^{*}_{i}n_{i}^{id}(T,\tilde{\mu}_{i})]^{-1}$
appears.
For example, the HM entropy density is given by
$$ s_{h}\;=\; \frac {\sum_{i} s_{i}^{id}(T,\tilde{\mu _{i}})}
{1+\sum_{i} V^{*}_{i}n_{i}^{id}(T,\tilde{\mu}_{i})}\;.$$

For the QGP we use the ``cut-off" model [10] which omits the
low-momentum contribution ($ k\le k_{c}$)
of the quark and gluon spectra.
In contrast to the bag model EOS for the QGP, the cut-off model
reproduces the lattice results reasonably well (see [11]). Besides,
it has one
more advantage in our problem. When we include supercooled states of
the QGP
we still have a positive QGP pressure whereas in the bag model we
would
encounter negative values of QGP pressure even for a small amount of
supercooling.

In our consideration
we have restricted ourselves to a pion-nucleon gas with the proper
volumes chosen as
$V^{*}_{\pi}=V^{*}_{N}=1.63\;\mbox{fm}^{3}$ (which gives a radius of
$0.73$~fm [9]) and have chosen $k_{c}=900$~MeV
(the same constant value for both $u,d$-quarks and gluons)
in order to get a reasonable $(T$--$\mu)$-phase diagram.
The phase transition line obtained by equating the HM and QGP
pressures is shown in Fig.2.

If we admit supercooled QGP states for the initial states in shock
transitions and/or superheated HM states for the final states, the
position of the TA in the $(\epsilon$--$p)$-plane
changes drastically. It does not
pass in this case
through the initial (non-equilibrium) point $(\epsilon _{o},p_{o})$.
The TA (7) for
our EOS and a supercooled initial QGP state lies in the following
regions of
the $(\epsilon$--$p)$-plane (see Fig.1):
I (s.l. deflagration),
II (unphysical region both for (6) and (10)),
III (t.l. detonation) and IV (s.l. detonation).
In Fig.3 we show several examples of TAs for the supercooled QGP
states
with $\mu_{o}=1000$~MeV and $T_{o}= 127.5, \;
124.0, \; 122.25$~MeV.
Black points correspond
to thermodynamically stable (entropy growth condition (3) is
fulfilled)
shock transitions to the final hadron states.
In a s.l. deflagration from QGP states, one observes the release of
the large
latent heat which is transformed into the collective motion of the
produced
HM (see Ref. [4]). A s.l. detonation of supercooled QGP (considered
in Refs. [5]) has quite different behaviour.
The QGP has a large velocity and small density. On the other
side of the shock front HM appears with lower velocity but larger
density. The space-time position of a discontinuity hypersurface was
not however considered in the previous studies of shock-like QGP
hadronization. We stress that this hypersurface is most probably
(see examples below) the
hypersurface with time-like normal vector. It leads to a t.l.
detonation
of supercooled QGP. Final HM states in this case belong to the
intermediate part of TA between s.l. deflagration and s.l. detonation
(see Figs.1 and 3).

The simplest solution of the t.l. shock equations (8,9) is
\begin{equation}
  \tilde{v}^{2}=\tilde
{v}_{o}^{2} = 0\;, \; \; \epsilon = \epsilon_{o}, \; n = n_{o}, \;\;
p \neq p_{o}\;.
\end{equation}
This is a time-like analog of the simplest
s.l. shock solution
$  v^{2}=v_{o}^{2} = 0,\; p = p_{o},\; \epsilon \neq \epsilon_{o},\;
n \neq n_{o}$
which is called a
contact discontinuity. Relations (14) take place, for example, in the
one-dimensional scaling expansion  (Bjorken model [12])
$v=x/t$ where the normal vector to the
hypersurface of constant proper time
$(t^{2} - x^{2})^{1/2} = const$
(in which the
thermodynamical parameters of the expanding QGP are fixed) has
the form $\Lambda^{\nu} = (1,0)$ in the rest frames of each fluid
element.
The minimal supercooling of the QGP to have a thermodynamically
stable t.l.
shock hadronization of this type is shown in Fig.2.
Note that for $\mu=1000$~MeV, the temperature for minimal
supercooling is $125.78$~MeV, so that the initial QGP state lies
above the minimal supercooling line for TAs in Fig.3a and below
in Figs.3b and 3c.

In the general case when $\tilde{v}_{o}^{2}$, defined by the
hydrodynamical
solution, is not zero, Eqs.(7,10) give us all the quantities of the
final state.

In conclusion we have studied energy-momentum and baryonic number
conservation in shock-like hadronization of a supercooled QGP. In the
expanding
system the space-time hypersurface of ``critical" (hadronization) QGP
parameters is most probably a hypersurface with a time-like normal 4-
vector.
This is the case for the Bjorken model in 1+1 dimensions and also for
all
known hydrodynamical solutions in the central rapidity region.
This means that the hadronization of a supercooled QGP arises as a
t.l.
shock which is different from the (standard) case of s.l. shock
hadronization.  Several examples of the TA have been analyzed in the
$(\epsilon $--$p)$-plane (Fig.3) to show different physical regions
for final
hadron states. The simplest t.l. shock hadronization (14) takes place
for the
scaling expansion of the QGP.
We have also found (see Fig.3) that solution (14) corresponds to the
minimal
supercooling for thermodynamically stable t.l. shock
hadronization. This minimal supercooling, as it is seen
from Fig.2, is not very strong
especially at small baryonic density.
For zero baryonic density the minimal
supercooling temperature is only 6\%
smaller than that of the phase transition.

\subsection*{Acknowledgements}
We acknowledge the financial support of the Foundation for
Research Development, Pretoria. M.I.G. is indebted to L.P.Csernai for
useful discussions of time-like shocks. He also gratefully
acknowledges
the warm hospitality at University of Pretoria,
where this work has been done.
\clearpage
\subsection*{Figure Captions}
{\bf Fig.1.} Possible final states in the
(energy density--pressure)-plane
for shock transitions from the initial state $(\epsilon_{o},p_{o})$.
I and IV are the physical regions for s.l. shocks, III and
VI for t.l.  shocks. II and V are unphysical regions for both types of
shocks. Note, that only states with $p \leq \epsilon$ are
possible for any physical equation of state in relativistic theory.
\\[0.3cm] {\bf Fig.2.} The phase diagram of  strongly interacting
matter in $(\mu$--$T)$-plane. The solid line shows the phase
transition between
HM and QGP.  The dashed line corresponds to the minimal supercooling
of
the QGP having thermodynamically stable t.l. shock hadronization (14)
in the
scaling expansion of the QGP.
\\[0.3cm] {\bf Fig.3a-c. } Taub adiabats for the
supercooled QGP initial states with $\mu_{o} =1000$~MeV and
$T_{o}=127.5$~MeV (a), $T_{o}=124.0$~MeV (b), $T_{o}=122.25$~MeV (c).

Black
points correspond to the thermodynamically stable (entropy growth
condition
(3) is fulfilled) shock transitions to the final hadron states. White
points
are thermodynamically unstable shock transitions.  The TA in the
unphysical
region II (where both $v_{o}^{2},v^{2}$ (6) and $\tilde{v}_{o}^{2},
\tilde{v}^{2}$ (10) are negative) are shown by dashed lines.
The points marked by large circles on the boundary between regions
III and IV
correspond to the limiting (unphysical) case when both t.l. shock
velocities (6) and s.l. (10) in both phases
are equal to 1. The straight solid line is $p=\epsilon -\epsilon
_{o}+p_{o}$.

\end{document}